\documentclass[11pt]{article}
\usepackage{epsf}
\usepackage{graphicx}        % standard LaTeX graphics tool

\def\ll{\label}

\def\r1{(\ref{$1})}

\def\ba{\begin{array}{c}}

\def\ea{\end{array}}

\def\l{\left}
\def\l({\left(}
\def\r){\right)}
\def\r{\right}

 \def\be{\begin{equation}}
\def\bc{\begin{center}}
\def\ec{\end{center}}
\def\bit{\begin{itemize}}
\def\eit{\end{itemize}}
\def\ee{\end{equation}}
\def\ed{\end{document}}
\def\bea{\begin{eqnarray}}
\def\eea{\end{eqnarray}}
\def\efr{\end{flushright}}

\begin{document}
\title{{ Exact non-circular symmetric  $N$-skyrmions in helical magnets 
without inversion symmetry  }}
%85
\author{ Anjan Kundu\\
 Theory Group \& CAMCS,
Saha Institute of Nuclear Physics\\
 Calcutta, INDIA\\
{anjan.kundu@saha.ac.in}} \maketitle
%\newpage
%\noindent Short Title: {\it Exact N-Skyrmions in magnets}
%\noindent {PACS:} 02.30.lk,
%integrable system02.30.jr,
%PDE
% 05.45.Yv,
%soliton11.10.Lm,
%nonlin nonloc FT
%\\
%{\it Key words}:\\
\vskip 1cm

%\begin{abstract}
\noindent {\bf Topological objects of intricate structures have been found in a wide
range of systems, like  cosmology to  liquid crystal$^{1}$, 
 DNA chains$^{2} $, superfluid $^3{\rm He}$ (Ref. 3),
 quantum Hall magnets$^{4,5}$, Bose-Einstein
condensates$^{6} $ etc.
   The topological spin texture in helical
magnets, observed  in  recently$^{7,8}$,
 makes a new entry into this fascinating phenomenon.
The unusual magnetic behaviour of
helimagnet MnSi,  noticed in recent years$^{9-12}$, prompted the suspicion that the
magnetic states arising in such crystals are of topological nature$^{13-15}$. 
Experiments$^{7,8}$  based on  the topological  Hall effect
 confirmed   such topologically nontrivial states
as  the skyrmions$^{16}$,
located on a plane$^{17,18}$ perpendicular to the applied magnetic field. 
% This new form of crystalline order is attributed to
%the lack of space-inversion symmetry, leading to the chiral spin interaction.
However, the available models close to MnSi,   
investigating the formation of 
  skyrmion  states$^{4,5,6,13,14} $, 
are based  mostly on  approximate or   numerical methods.
%111
We present here a theoretical model for chiral  magnets
 with competing exchange and
Dzyaloshinskii-Moriya  type interaction, which 
leads to an exact skyrmionic solution with integer topological charge $N$.
Such topologically stable spin states with analytic solution
 on a two-dimensional  plane$^{19}$ show  helical
 structures of partial order, without  
 inversion and circular symmetry. These  
 exact  N-skyrmions, though represent higher excited states,
correspond to  the  lowest energy stable configuration in each topological sector 
 and are likely to appear in  MnSi  under suitable experimental conditions. The present exact topological
solitons, with explicit non-circular symmetry  could   be applicable also
to other fields, where skyrmions are  observed, especially 
in natural systems with less symmetries.
}
%111+107=227

%\end{abstract}

Topological properties can be  revealed through a mapping from a continuum 
space to a differentiable manifold. Therefore for describing topological
objects one has to shift from the lattice to a continuum picture.
In a magnetic model with the spin configuration  varying slowly over the
lattice spacing,  one can approximate using  a  long 
   wavelength  description, 
  the $d $  dimensional lattice   to a continuum space $R^d$.
At the same time  the associated 
spin ${\bf S}_{\bf j} $   would go to a vector field $ n^a({\bf x }), a=1,2,3 $ of
unit length $|{\bf n} |^2=1 $, at the classical limit, after a  proper 
renormalisation$^{20}$, which might induce  topological invariants of
 different nature, at different   dimensions  $d , $
 under suitable boundary conditions  $^{20-23}$. 

We intend to use this construction 
on a two-dimensional $xy $-plane perpendicular to the applied magnetic field,
to simulate the result on MnSi found through the 
topological quantum Hall experiment.
 The physically motivated 
boundary condition  demands   that 
at large distances: $|{\bf x}|=\rho  \to \infty $,
 the spin field  $n^a({\bf x})$  should go to its vacuum solution, orienting itself
 to a fixed vector  $n^a_\infty=\delta_{a3} $,  along the applied field. 
This condition in turn 
   introduces  a  nontrivial topology by     
identifying  the infinities of the coordinate  space to a single point, which
  compactifies the  vector space  $R^2$ to a sphere  $ S^2_{{\bf x}} $
 and   defines  thus a mapping:
 $ S^2_{{\bf x}} \to S^2_{{\bf n}}$, 
linked to the  nontrivial  homotopy group  $ \pi _2(S^2)=Z  $
 (see Fig. 1). Each topological 
sector is labeled 
 by an integer $N $-valued  topological charge 
$Q= \int d^2x \Phi({\bf x})$,
which  defined as  the degree of this mapping,
  shows how many times such a skyrmionic 
 field configuration ${\bf n}({\bf x})$
winds up  or covers the  target sphere, when
 the coordinate space is swept  
once. The corresponding charge
  density can be   
expressed through the  spin vector field as
\be 
  \Phi({\bf x})=\frac 1 {8 \pi}( {\bf n} [\partial_x{\bf n}
 \times \partial_y {\bf n}] )  . \ll{Phi}\ee

\begin{figure}[h!]
\qquad \qquad \qquad \qquad \qquad \framebox{
\includegraphics[width=8.2cm,height=4.1 cm]{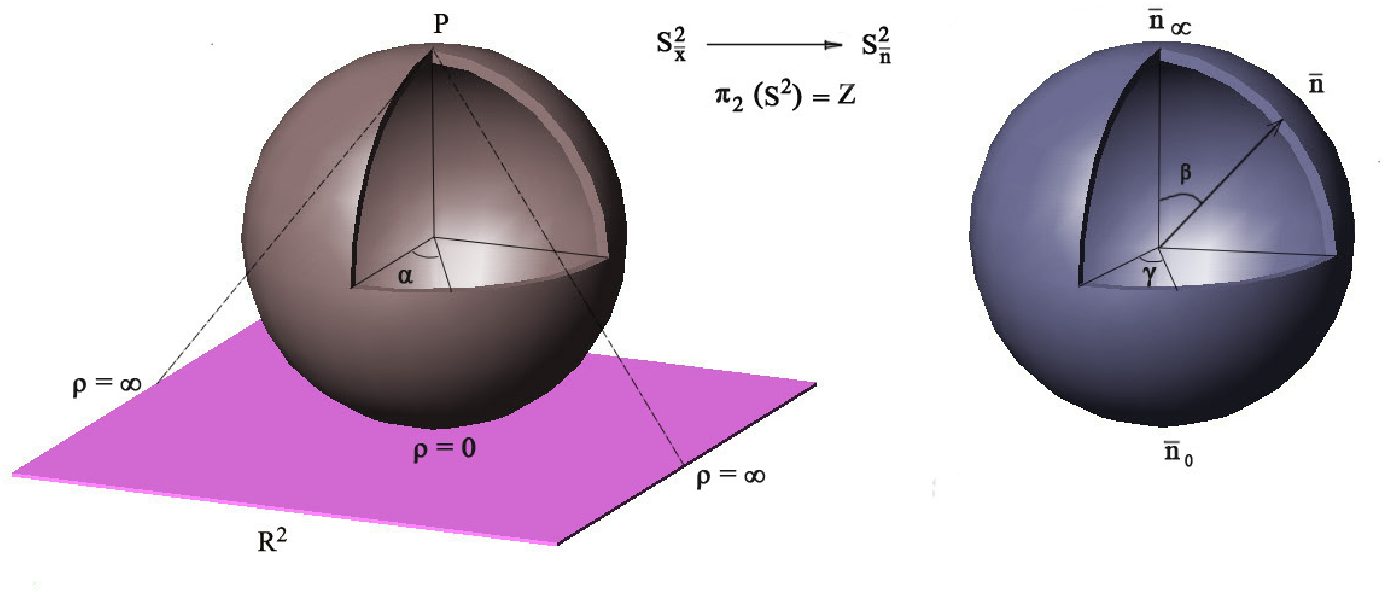}
%{circles2f2.eps}
% \quad \quad \quad \quad \quad \quad \quad \quad 
%\includegraphics[width=5.cm,height=4.1 cm]{n121.eps}
} 
%\caption{
%\end{center}
\end{figure}
\noindent  {\bf Figure 1 $|$ The induction of nontrivial topology}.
 The two-dimensional  coordinate space ${\bf x} \in R^2 $,
 due to the boundary condition 
$ \lim_{\rho \to \infty}{\bf n}={\bf n}_\infty =(0,0,1), $  compactifies to a
 2-sphere $ S^2_{{\bf x}}$, with its north-pole $P$ identified to all points at
the space-infinities. The unit vector field ${\bf n}({\bf x}) \in S^2_{{\bf
n}}$,  therefore describes a sphere to sphere mapping with the integer-valued  
topological charge $Q=N$, defined as the degree or the winding number  of this mapping.
\vskip .5cm 

%}

Another important idea, that we  use for constructing  our magnetic model is to 
maintain a lower 
bound \ $H \ge
const. |Q| $ \  for   energy  $H$ of the system   through 
its topological charge $ Q$. 
Note  that, such an  energy bound would guarantee 
the stability of the finite energy solution in each
topological sector, since the topological charge is conserved 
independent of the dynamics of the system and can not be changed 
by unwinding the spin configuration by smooth transformations.
Moreover, when   the energy reaches its lowest value saturating this  bound:
 $H=H_{min}=const. |Q| $,  called  Bogomolny limit$^{24}$, some extraordinary
thing happens in almost all known  models $^{17,18,21,23-25} $:  
it yields an exact solution to the field model, in some cases for all
topological
sectors.

The unusual properties of the helical magnets like MnSi and especially their 
partial helical order created by  the  skyrmion spin states
 are believed to be due 
 to the competing  forces between the ferromagnetic  
 exchange and an effective Dzyaloshinskii-Moriya  (DM) type interaction. The ferromagnetic interaction 
 tries to align the spins parallel to each other,
 while the DM-term with the broken inversion symmetry tends to 
orient them perpendicular to each other, settling finally to a 
partial helical order of topological origin. 
 Moreover, the skyrmionic spin texture  in MnSi,
 as observed in the recent  topological    Hall 
experiments,
  appears  in the plane perpendicular to the applied 
  magnetic field.
Therefore to model such a system we construct our
 Hamiltonian $H$ from a  standard 
Heisenberg  ferromagnet in  two-dimensions:
$H_0=-J\sum_{<{\bf j}, {\bf j'}>} {\bf S_j}{\bf S_{j'}}, $
where the  spin ${\bf S_j} $ , located at
  site ${\bf j}= (i,j)$ interact with its   nearest-neighbour at
  $ {\bf j'}= (i, j\pm 1) . (i\pm 1, j)$, in the 2-dimensional lattice. 
We include next the  DM term:
$H_{DM}=J_1\sum_{<{\bf j}, {\bf j'}>} ({\bf D}_{{\bf j}, {\bf j'}} \cdot [{\bf S}_{\bf
j}\times {\bf S}_{\bf j'}]) ,$  
which in spite of its  traditional form, 
bears   some  crucial differences in the structure  of its
 Dzyaloshinskii (D)-vector ${\bf D}_{{\bf j}, {\bf j'}}$,
 which would result to our
 field model allowing exact  skyrmion solution. Firstly, we take
  the D-vector aligned  along the  applied field,
which leaves nontrivial  only  its third-component  ${D}^3_{{\bf j}, {\bf j'}} $,
breaking the rotational symmetry in the internal spin-space.  
Our  next   deciding  assumption
 is the nonlinear nature of this vector,
given through a spin-dependent structure:  
 ${D}^3_{{\bf j}, {\bf j'}}={D}^3(S^3_{\bf j}, S^3_{\bf j'})=f(S^3_{\bf j})
(S^3_{\bf j}-
S^3_{\bf j'})$. Note that the second factor   obeys  the 
 usual
antisymmetric exchange   of the D-vector, while the first factor, 
which we take in the explicit form $f(S^3_{\bf j})=(1-(S^3_{\bf j})^2)^{-1} $
manifestly breaks  the space-inversion symmetry of the vector:
 ${D}^3_{{\bf j'}, {\bf j}} \neq {D}^3_{{\bf j}, {\bf j'}}, $   
inducing   the same property  to the Hamiltonian
 of the system as  usual through the DM interaction.

 Our total Hamiltonian, which has  
the required broken inversion-symmetry and the competing spin interactions 
of opposite trends,  should   now be tested for its topological properties, for
which we have to go for   the  continuum  limit ${\bf S}_{(i,j)}\to 
{\bf n} ({ x, y})  $.
For constructing the corresponding  field model from the    2-dimensional
lattice model, we have to  take carefully the 
vanishing limit of the lattice constants    in both the lattice
 directions $(i,j) $, which would result, for example,
$ \ {\bf S}_{(i\pm 1,j)}-{\bf S}_{(i,j)}\to \pm \partial_x {\bf n}_{(x,y)},
\ \ {\bf S}_{(i,j \pm 1)}-{\bf S}_{(i,j)}\to \pm \partial_y {\bf n}_{(x,y)}
\ $  etc. This procedure 
 reduces our  
lattice spin-model $H=H_{0}+H_{DM} $ to a field  Hamiltonian with  
\be H_0=J\int d^2x (\nabla {\bf n})^2  \quad \quad  \mbox{and} \quad  \quad 
 H_{DM}=J_1\int d^2x \ f(n^3) \nabla  n^3 \cdot
 (n^1 \nabla  n^2- n^2 \nabla  n^1)    . \ll{H01} \ee
It is  not difficult to see that, at this continuum limit 
the chirality of three non-coplanar spins $\chi={\bf S}_{\bf j}\cdot [
{\bf S}_{\bf j'} \times {\bf S}_{\bf j''}] $ gets linked   to the 
 topological charge density   $\Phi({\bf x}) $,  expressed  
through the
${\bf n} $-field as (\ref{Phi}).

It is important to note,  that by  tuning  the coupling constants 
$J \to J_1 $, our field Hamiltonian $H$, as can be shown by  using a simple 
school-level geometric inequality$^{19}$,   would  acquire the  crucial lower bound 
: $H\geq  4 \pi \sqrt 3 J |Q| $, through the topological charge $Q $, the
significance of which for the topological stability of the 
	 solutions is emphasized above.
Focusing now on  the conjecture, that the Bogomolny limit, at which
the energy bound is saturated, should yield the
exact finite-energy soliton solution,   we
  look for the situation  when this limit could be reached 
 for   nontrivial contributions 
from both  parts of the Hamiltonian. We find   that 
this is possible  for a field configuration $ {\bf n}^* $,
 where the space-inversion as well as the circular symmetry is 
lost.
%, where $\beta= \beta(\rho,\alpha), \gamma = \gamma (\rho,\alpha) $
% are the angles describing the
%field on the sphere and $(\rho,\alpha) $ are the polar coordinated on the
%plane
 More precisely, the spin texture ${\bf n}^*=(\sin \beta \cos \gamma,
\sin \beta \sin \gamma, \cos \beta ), $ should be    such that,  
  the directions
 $\nabla \beta $ and $\nabla \gamma $, defining the helical order, 
would cross each other   at an angle $60^o $,  with 
three field vectors: ${\bf a}= \nabla \beta $ , ${\bf b}=\sin \beta \nabla \gamma$
and ${\bf c}={\bf a}-{\bf b} , $  forming an
isosceles triangle. Under this special geometric condition,
  the energy bound is exactly
 saturated attaining its 
lowest value $H_{min}[{\bf n}^*]=4 J \pi \sqrt 3 N , $ in  each sector with
$Q=\pm N $.
Analysing this intriguing geometry$^{19}$ 
 we can extract the exact magnetic field solution   as 
\be  \sin \beta= 2 (\rho \rho_0)^{\frac {N \sqrt 3} {2}}e^{\frac \alpha 2 N }
/((\rho )^{ {N \sqrt 3} }+(\rho_0)^{{N \sqrt 3} }e^{ \alpha  N }) \ \mbox{
and} \ , \ 
\gamma =\pm N \alpha ,\ll{n*}\ee 
in  polar coordinates $(\rho, \alpha) $ with  a constant 
scaling parameter  $\rho_0 $. 
This exact  solution, which can be checked by direct insertion, 
obeys the required boundary condition 
$ {\bf n}(\rho \to \infty)={\bf n}_\infty =(0,0,1),  $
 and yields the topological charge
$Q=\pm N$. The distribution  of the charge density $\Phi(\rho, \alpha) $,
  which  coincides also
with the energy density at the solution manifold ${\bf n}^* $
  is shown in Fig. 2,  together 
with the magnetization $ \sin \beta (\rho, \alpha)$   and the spin component
$n^3=\cos \beta (\rho, \alpha) $ 
% $m=\sqrt((n^1)^2+(n^2)^2)=\sin \beta $
, calculated for the exact result (\ref{n*}), 
specific to the first excited  state with $Q=1 $.
The figure shows that at the skyrmion axis $\rho =0, $
 the charge density becomes maximum, vanishing  gradually with the distance,
 while the magnetisation and the spin component exhibit   a toroidal structure
 without
 circular and  space-inversion symmetry. Their maximum/minimum is attained 
 at a distance $ \rho_m= \rho_0 \ e^{\frac {\alpha} {\sqrt {3}} }$, varying
 with $\alpha $, which describes   an 
 intriguing form of the torus, with its   radius  spreading out with 
  increasing  polar angle.  
As  revealed in  quantum Hall experiments$^{7,8}$,  
the peculiarly large Hall conductance  
%induced  by the spin chirality, 
is created  
by an effective magnetic
field proportional  to the topological charge density.
Therefore the detailed and exact form of the 
charge density together with the magnetization and the spin field component,
 found here 
based on the exact skyrmion solution, should serve as a guiding resource 
for the  future precision  experiment.  
%Fig2ab
\newpage

\begin{figure}[h!]
  \quad \quad \framebox{
\includegraphics[width=7.2cm,height=5.1 cm]{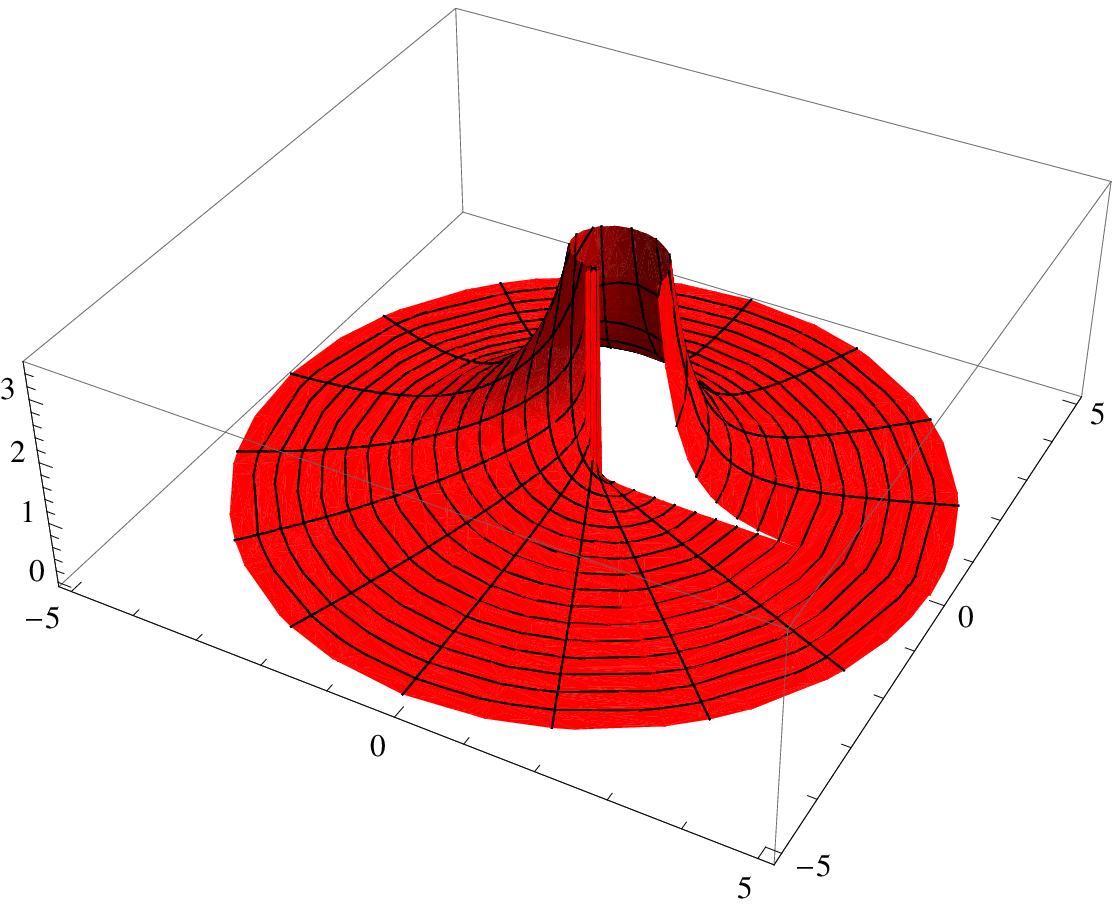}
%{Energd1.eps}
}
\end{figure}
\qquad \ \  \qquad \qquad  \qquad \ \ {\bf  a} 

\begin{figure}[h!]
 \quad \quad \framebox{  
\includegraphics[width=7.2cm,height=5.1 cm]{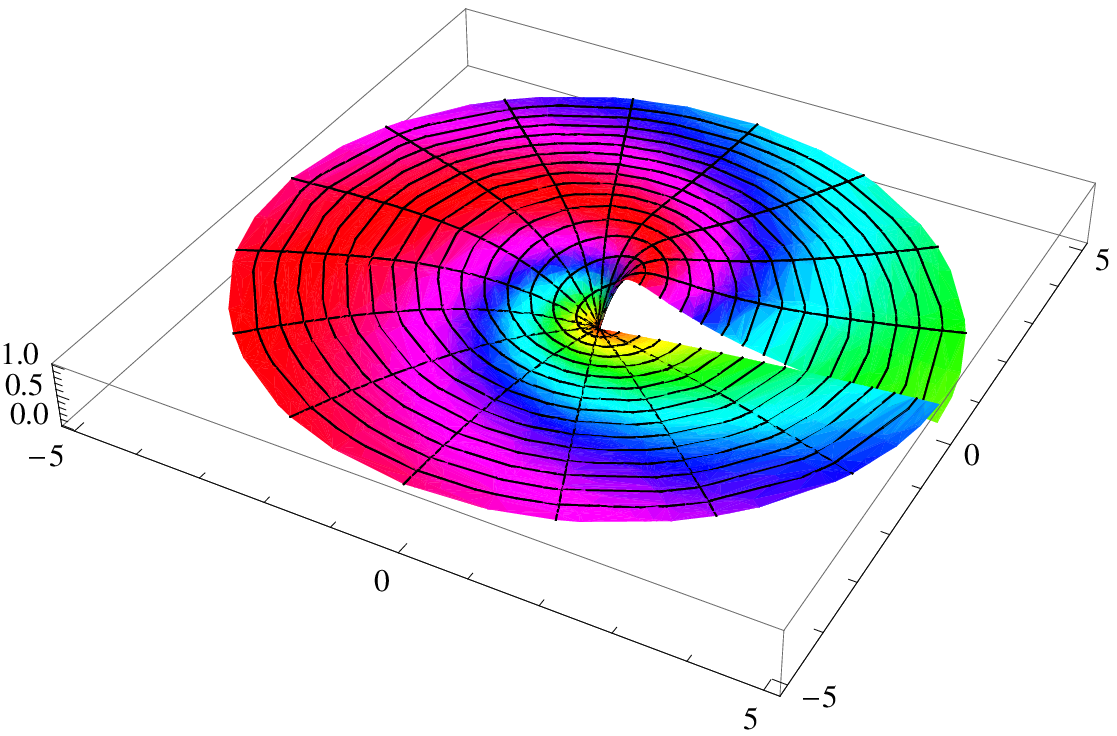}
%{n121.eps}
}\quad   \qquad \framebox {\includegraphics[width=7.2cm,height=5.1cm]{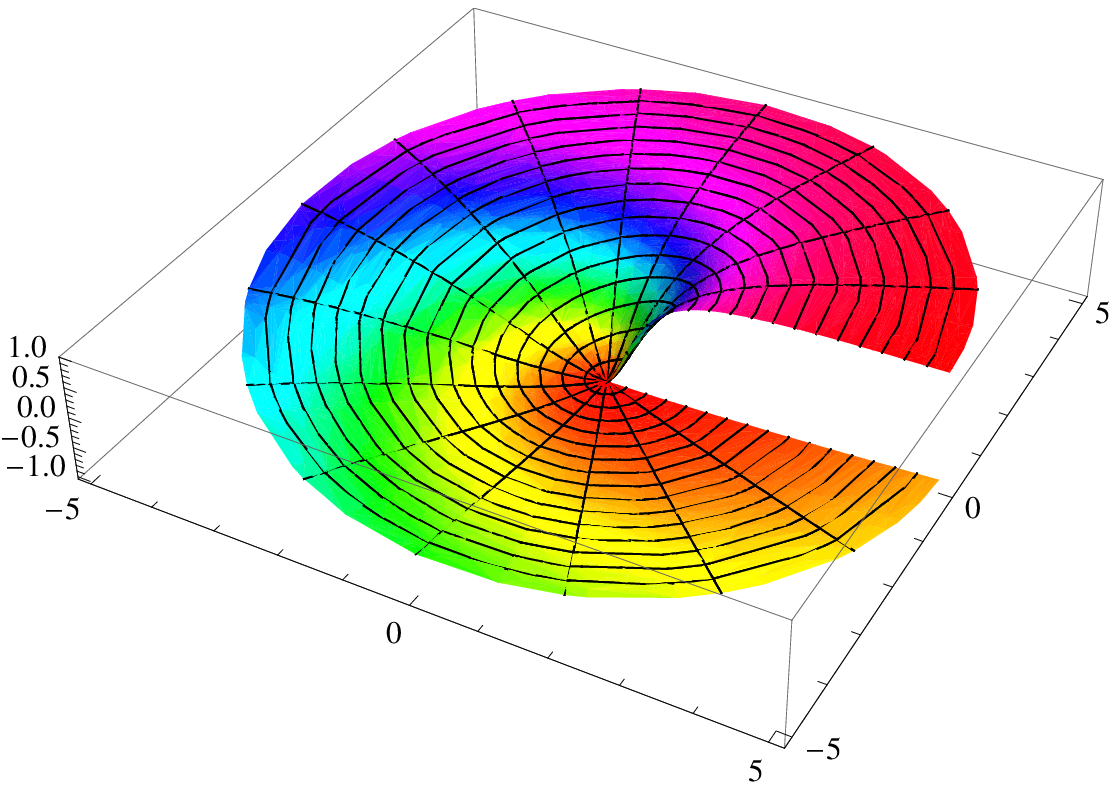}
%{n31.eps}
}
 %Fig2ab
%\caption{ 
%}
%\end{center}
\end{figure}
 \qquad
 \qquad \qquad  \qquad \  \quad  \ {\bf b}  \qquad \qquad  \qquad  \ \ \qquad \ \
  \ \   \ \ \ \qquad \ \qquad \qquad \qquad \qquad
 {\bf c }

\noindent {\bf Figure 2 $|$ Exact skyrmion state with $Q=1 $}. 
 {\bf a,} The charge density $\Phi(\rho, \alpha)=\sin ^2 \beta(\rho, \alpha) /\rho^2 .$
  which  coincides also with the energy density 
for the exact skyrmion. {\bf b,} The magnetization $|{\bf
m}|=((n^1)^2+(n^2)^2)^{\frac 1 2}=\sin \beta (\rho, \alpha),  $ and {\bf c,}
 Spin component $
n^3=\cos \beta (\rho, \alpha)$,   
showing toroidal structure with explicitly broken  inversion and circular
 symmetry.

\vskip .5cm

The graphical result based on the numerical calculation on
a model for $MnSi ,$   proposed in   Ref 13, shows that 
   the skyrmion lattice  derived as a spontaneous
ground state
solution,  preserves
 the circular as well as  the space-inversion symmetry, when projected on  two-dimensions.
However, our exact result (\ref{n*}) based on our   magnetic model with
 the broken space-inversion and 
the spin rotational symmetry, induced    by    the  DM interaction,
gives a different picture. It  shows, as depicted in 
Fig. 3, that 
the
partial helical order of the spin texture on the $xy$-plane manifestly
 breaks both the circular and  
the inversion symmetry.
 The future
experiments  capable of identifying the finer 
details of the topological spin texture 
 should  be able to settle this
issue.  At the theoretical level, it is  significant to note also   that, the
exact  topological  soliton  
without   circular symmetry,  presented here,  is a nontrivial  generalisation over
the well known  solution of the nonlinear $ \sigma$-model$^{17,18,21}$,
 exhibiting this symmetry. In fact
all other known exact solutions of the field models$^{23-25} $, obtained at the Bogomolny
limit,  
are basically spherical  symmetric, except only the present one.

Another valuable information, e.g. the 
 detail spin structure in higher topological sectors with arbitrary integer
charge
 $Q=N \geq
1 $, can be found easily from our exact solution (\ref{n*}). 
Although the solution  with higher topological charge was investigated 
by  using a combination of analytic and
numerical methods   in
connection with the Hall  skyrmions $^{5}$ and recently for the bilayer graphine$^{26}$,
 no such study, as far as we know, 
 is undertaken   for
the
magnetic models. Therefore the present result, 
designed specifically for the magnets like MnSi.
and more significantly having  purely
analytic  result  with manifestly broken 
inversion  symmetry,  is novel in many respect.
 Such an exact skyrmion  
 with higher topological charge $Q=3$,  is shown in Fig. 3b. These stable states in higher
sectors with more intricate texture, as we see in the figure, would be
challenging to detect in topological Hall experiments. 

The topological soliton  with non-circular symmetry, presented here,
is exceptional  as an exact field theoretic solution and would be
interesting for its detection and applications in other systems, lacking
such a symmetry.
\vskip .5cm

%\newpage
%Fig3ab
\begin{figure}[h!]
\qquad \ \ \framebox{
\includegraphics[width=7.2cm,height=6.1 cm]{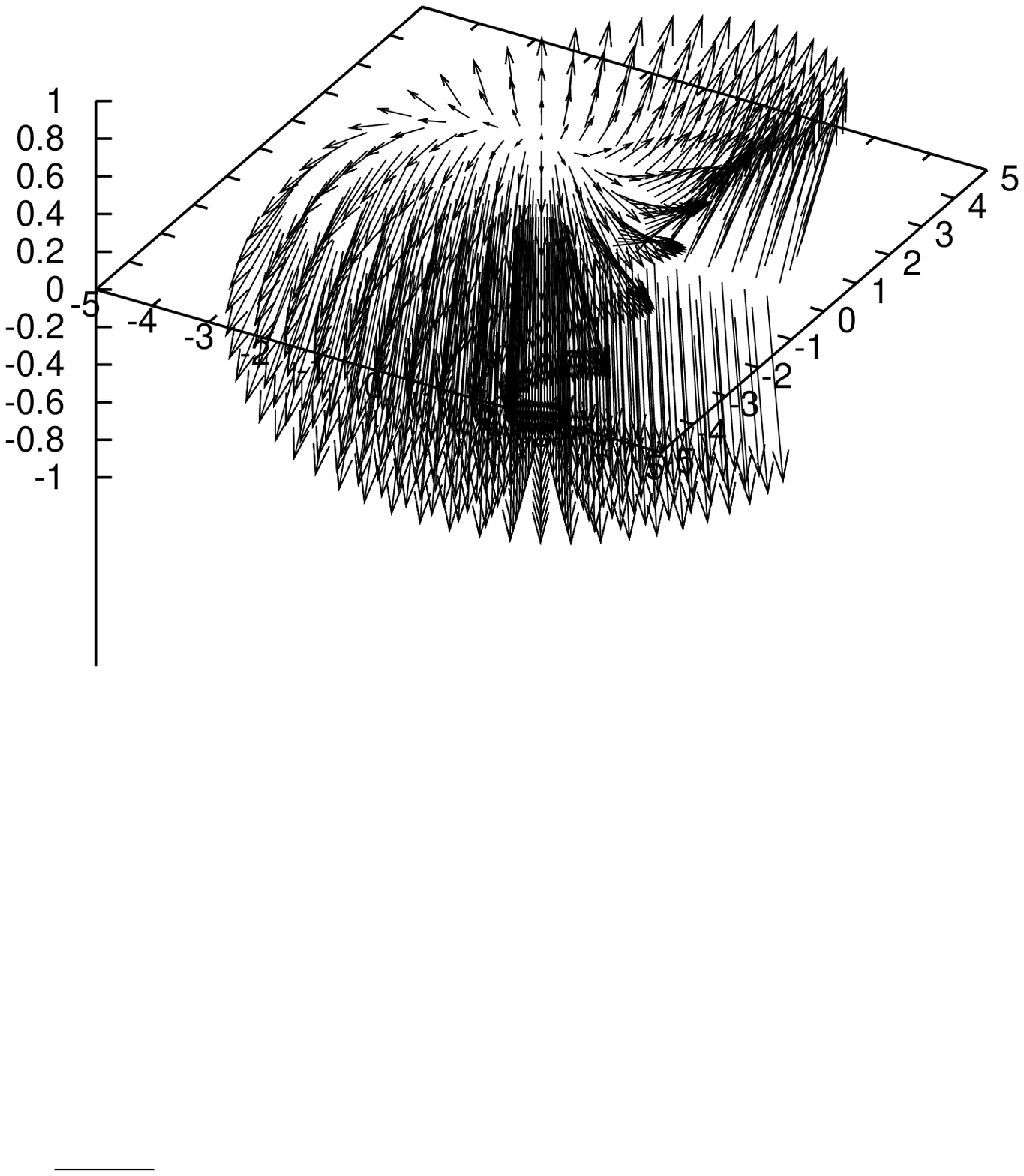}
%{Vn10.eps}
}   \quad \ \ \quad \framebox{  
\includegraphics[width=7.2cm,height=6.1 cm]{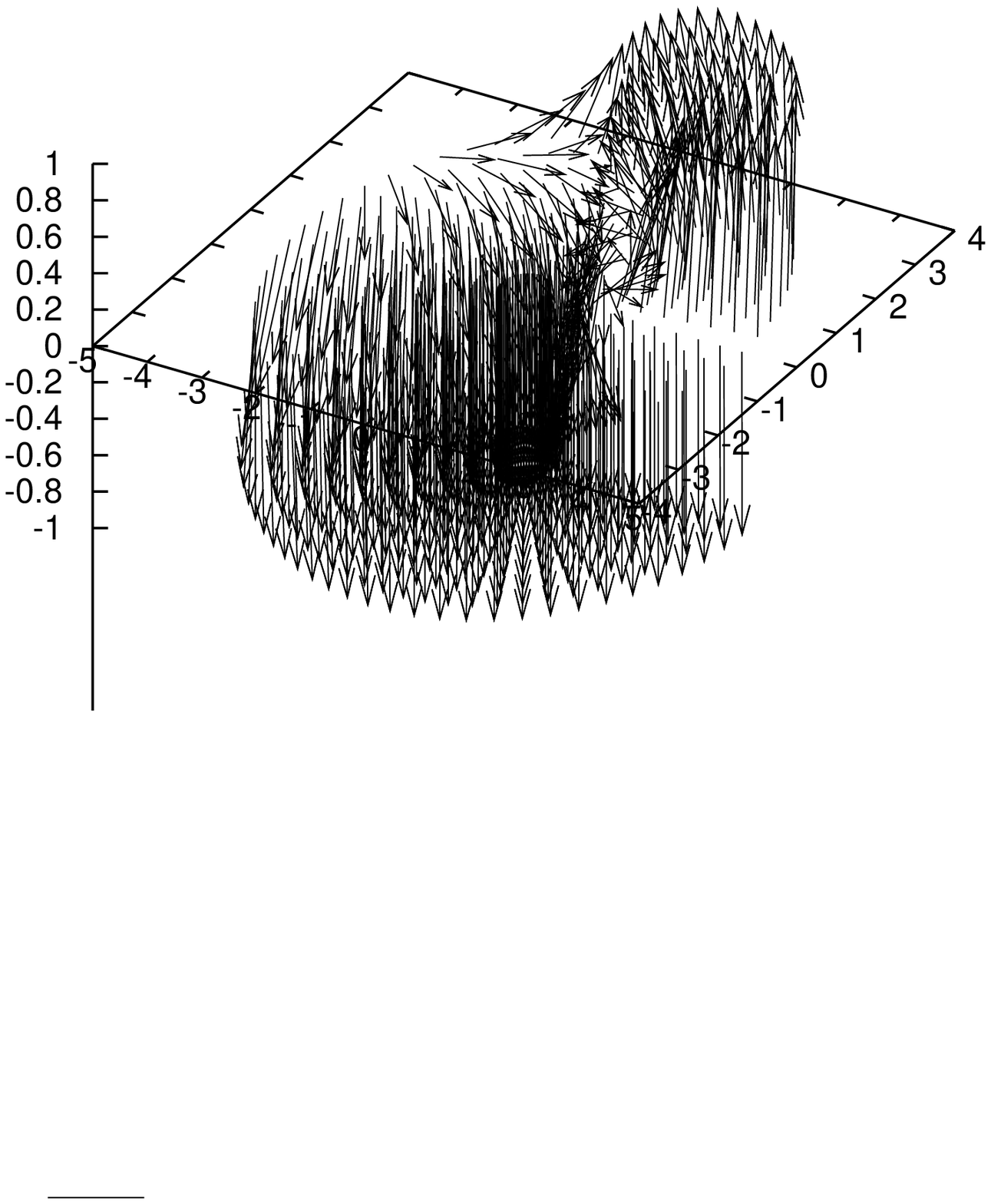}
%{Vn30.eps}
}
 %Fig2ab

%\caption{
%\end{center}
\end{figure}
\qquad \ \  \ \ \qquad \qquad \   \qquad \ \ {\bf  a} \qquad \qquad \qquad \qquad \qquad  \qquad
\qquad  \qquad \qquad 
\ \ \ \ \  \ \ \ \ {\bf b}\\
\noindent  {\bf Figure 3 $|$  Skyrmion spin texture with partial  helical order,
 described by the   exact solution of the spin vector field ${\bf n}({\bf
x}). \ 
$}{\bf a , } \  with
the topological charge  $Q=1$ and 
 {\bf b, } \    $Q=3$.  Both the  figures demonstrate manifestly broken space-inversion and the
circular symmetry on the $xy$-plane, with a distinct   difference  in their
  helical pattern,  describing different    degrees of  mapping in these
two cases.
%}

 \end{document}